\documentclass{emulateapj}
\usepackage{epsfig}
\usepackage{apjfonts}

\newcommand{\ltsim}{\protect\raisebox{-0.5ex}{$\:\stackrel{\textstyle <}
	{\sim}\:$}}
\newcommand{\gtsim}{\protect\raisebox{-0.5ex}{$\:\stackrel{\textstyle >}
	{\sim}\:$}}

\newcommand{\rb}{r_{\rm B}}

\newcommand{\mdotbh}{\dot{M}_{\rm BH}}
\newcommand{\mdotturb}{\dot{M}_{\rm turb}}
\newcommand{\mdotomega}{\dot{M}_{\omega}}
\newcommand{\mdotzero}{\dot{M}_{0}}

\newcommand{\rbh}{r_{\rm BH}}
\newcommand{\tbh}{t_{\rm BH}}

\newcommand{\calm}{\mathcal{M}}
\newcommand{\oms}{\omega_{*}}
\newcommand{\phimean}{\phi_{\rm mean}}
\newcommand{\phimed}{\phi_{\rm med}}
\newcommand{\smdot}{\sigma_{\dot{M}}}
\newcommand{\cs}{c_{\rm s}}
\newcommand{\msun}{M_{\odot}}
\newcommand{\sr}{\sigma_{\rho}}

\begin{document}

\submitted{Accepted October 13, 2005}

\title{Bondi-Hoyle Accretion in a Turbulent Medium}

\author{Mark R. Krumholz\footnote{Hubble Fellow}}
\affil{Department of Astrophysical Sciences, Princeton University, Peyton
Hall, Ivy Lane, Princeton, NJ 08544-1001}
\email{krumholz@astro.princeton.edu}

\author{Christopher F. McKee}
\affil{Departments of Physics and Astronomy, University of California,
Berkeley, Berkeley, CA 94720}
\email{cmckee@astron.berkeley.edu}

\author{Richard I. Klein}
\affil{Astronomy Department, University of California, Berkeley,
Berkeley, CA 94720, and Lawrence Livermore National Laboratory,
P.O. Box 808, L-23, Livermore, CA 94550}
\email{klein@astron.berkeley.edu}

\begin{abstract}
The Bondi-Hoyle formula gives the approximate accretion rate onto a
point particle accreting from a uniform medium. However, in many
situations accretion onto point particles occurs from media that are
turbulent rather than uniform. In this paper, we give an approximate
solution to the problem of a point particle accreting from an ambient
medium of supersonically turbulent gas. Accretion in such media is
bimodal, at some points resembling classical Bondi-Hoyle flow, and in
other cases being closer to the vorticity-dominated accretion flows
recently studied by Krumholz, McKee, \& Klein. Based on this
observation, we develop a theoretical prediction for the accretion
rate, and confirm that our predictions are highly consistent with the
results of numerical simulations. The distribution
of accretion rates is lognormal, and the mean accretion rate in
supersonically turbulent gas can be substantially enhanced above the
value that would be predicted by a naive application of the
Bondi-Hoyle formula. However, it can also be suppressed by the
vorticity, just as Krumholz, McKee, \& Klein found for non-supersonic
vorticity-dominated flows. Magnetic fields, which we have not included
in these models, may further inhibit accretion. Our results have
significant implications for a number astrophysical problems, ranging
from star formation to the black holes in galactic centers. In
particular, there are likely to be significant errors in results that
assume that accretion from turbulent media occurs at the unmodified
Bondi-Hoyle rate, or that are based on simulations that do not resolve
the Bondi-Hoyle radius of accreting objects.
\end{abstract}


\section{Introduction}

Accretion of gas from a background medium onto a small particle occurs
throughout astrophysics. Examples range from protostars accreting from
their natal clouds to black holes accreting interstellar gas during
galaxy mergers. In such cases, one wishes to know the rate of mass
accretion onto the point-like object. Bondi, Hoyle, and Lyttleton
\citep{hoyle39,bondi52} derived the classic
solution to this problem when the background gas is uniform and either
stationary or moving with constant velocity relative to the accreting
object. More recent numerical work has demonstrated that their
treatment, while missing some instabilities that cause the accretion
rate to be time-dependent when the background gas is moving rapidly
relative to the accretor, gives essentially the correct accretion rate
\citep{fryxell88, ruffert94a, ruffert94b, ruffert97, ruffert99,
foglizzo99}. The accretion rate is reasonably well-fit by
\citep{ruffert94b}
\begin{equation}
\label{bhformula1}
\mdotbh = 4 \pi \rho_{\infty} G^2 M^2 c_{\infty}^{-3} \left[
\frac{\lambda^2 + \calm^2}{\left(1+\calm^2\right)^4}
\right]^{1/2},
\end{equation}
where $\rho_{\infty}$, $c_{\infty}$, and $\calm$ are the density,
sound speed, and Mach number of the flow far from the accreting
object, and $\lambda$ is a factor of order unity that depends on
$\calm$ and the equation of state of the gas. For $\calm=0$ (Bondi
accretion) in an isothermal medium, $\lambda=\exp(1.5)/4\approx 1.1$,
and we adopt this value in our analysis below.

However, many sources of gas accretion are supersonically turbulent,
and thus are not easily characterized by a constant background density
or velocity. A seed protostar accreting gas to which it is not bound
in a process of ``competitive accretion'' \citep[e.g.][]{bonnell97,
klessen00a, klessen01, bonnell01a, bonnell01b, bonnell04} inside a
turbulent molecular core is an example of such a
phenomenon. \citet{padoan05} suggest that the observed accretion rates
onto T Tauri stars may be the result of accretion from the remnant
molecular gas of the star-forming region.

Despite the ubiquity of the phenomenon, theoretical and numerical work
to date has not clearly shown how to extend the Bondi-Hoyle-Lyttleton
result to the case of a turbulent background medium. As a zeroth order
estimate, one might simply use the mean density and gas velocity
dispersion of the turbulent medium in the \citet{bondi52} formula in
place of $\rho_{\infty}$ and $c_{\infty}$. This would produce a
zeroth-order estimate of the accretion rate
\begin{equation}
\label{mdotzerodef}
\mdotzero \equiv 4\pi\bar{\rho}\frac{(G M)^2}{(\calm_0 \cs)^3},
\end{equation}
where $\overline{\rho}$ is the mean density, $\calm_0$ is the Mach
number of the turbulent flow, $\cs$ is the sound speed, and $M$ is
the mass of the accreting object. Of course this approximation is only 
valid for $\calm\gtsim 1$. \citet{padoan05}
suggest a somewhat more sophisticated approach in which one computes
the accretion rate by applying the Bondi-Hoyle formula at every point
within a turbulent medium and then taking the volume average.

However, numerical and analytic studies have shown that even a small
amount of vorticity in the accreting medium can substantially change
the accretion rate \citep{sparke80, sparke82, abramowicz81, fryxell88,
ruffert97, ruffert99, igumenshchev00a, igumenshchev00b, proga03,
krumholz05b} in regions where the overall flow velocity is small, and
in a supersonically turbulent medium the baroclinic instability is
likely to generate significant
vorticity. \citet{krumholz05b} use simulations and analytic arguments
to derive a formula analogous to the Bondi-Hoyle formula accretion in
a medium with vorticity, and application of this formula to turbulent
media suggests that the vorticity of a turbulent gas may be at least
as important as its velocity in inhibiting accretion. Magnetic fields, 
which we do not include here, may further reduce the accretion rate.

In this paper we derive an estimate for the accretion rate onto a
point particle in a turbulent medium, thereby extending the
Bondi-Hoyle-Lyttleton solution to this case. We limit ourselves to the 
case of an accretor substantially smaller than the medium from which
it is accreting, where the self-gravity of the medium is negligible in 
comparison to the gravity of the accretor, and where the accreting
medium is dominated by supersonic, isothermal turbulence. Stars
accreting in molecular clouds satisfy these conditions due to the
short times the gas requires to reach radiative equilibrium
\citep{vazquezsemadeni00}, and below we discuss other astrophysical
situations to which our resuls apply. In \S~\ref{theory} we propose a
simple  method for determining the accretion rate in a turbulent
medium as a function of the properties of the gas. In \S~\ref{sim}
we describe numerical simulations we have conducted to test our
theoretical model, and show that it provides an excellent fit. In
contrast, earlier proposed models give far less accurate
predictions. In \S~\ref{extension} we use our model to predict the
rate of accretion onto point particles in a turbulent medium as a
function of a few simple properties of the accreting gas, and in
\S~\ref{bhtdiscussion} we discuss the implications and limitations of our
approach. Finally, in \S~\ref{bhtconclusion} we summarize and present
our conclusions.

\section{A Simple Model for Accretion in a Turbulent Flow}
\label{theory}

In this section, we describe a simple \textit{ansatz} to relate the
rate of accretion onto a point particle in a turbulent gas to the
properties of the gas.
Since a turbulent medium has fluctuations in gas properties in space
and time, it is most convenient to characterize the accretion rate for
a particle placed within it using a cumulative probability
distribution function (PDF), which specifies the probability
$P_{\dot{M}}(<\dot{M})$
that a particle placed at a random position within the turbulent gas
will accrete at a rate less than $\dot{M}$. We wish to predict this
function in terms of the properties of the turbulent
gas. (Throughout this paper, we use $P$ to refer to
cumulative distribution functions and $dp$ to refer to the
corresponding differential distribution, or probability density,
functions.)

Our \textit{ansatz} is that we can roughly divide accretion in a
turbulent medium into two modes. In regions with relatively high
velocities and small vorticities, the vorticity should have little
effect on the flow pattern, and flow should resemble ordinary
Bondi-Hoyle accretion. For example, \citet{ruffert97, ruffert99} show
that in simulations with relatively little vorticity, the overall
accretion rate and flow pattern are quite similar to what one finds
with no vorticity. In regions where the flow velocity is relatively
small and the vorticity is relatively large, the small overall
velocity of the gas relative to the particle should have little
effect. Instead, the vorticity should dominate the flow, and the flow
pattern and accretion rate should resemble those found by
\citet{krumholz05b}.

Which mode occurs for a given particle will be determined by which one
would produce a lower accretion rate. Thus, our suggested procedure
for estimating $P_{\dot{M}}(<\dot{M})$ in terms of the properties of
the turbulent gas is to compute the function 
\begin{equation}
\label{bhturbformula}
\mdotturb(\mathbf{x}) \approx
\left[\mdotbh(\mathbf{x})^{-2} +
\mdotomega(\mathbf{x})^{-2}\right]^{-1/2},
\end{equation}
at every point $\mathbf{x}$ within the flow. Here, $\mdotbh$ is to be
computed with the Bondi-Hoyle formula (equation \ref{bhformula1}), using
the density $\rho(\mathbf{x})$ and velocity $v(\mathbf{x})$ at
$\mathbf{x}$ for $\rho_{\infty}$ and $v_{\infty}$, and using the
constant isothermal sound speed $\cs$. The quantity
$\mdotomega$ is the accretion rate in a vorticity-dominated medium. It
is a function of the density and vorticity
$\omega=\left|\nabla\times\mathbf{v}\right|$, given by
\citep{krumholz05b}
\begin{equation}
\label{mdotomegaformula}
\mdotomega=4\pi\rho_{\infty} \frac{(G M)^2}{\cs^3}\, 0.34\, f(\oms),
\end{equation}
where
\begin{equation}
\oms \equiv \omega \frac{\rb}{\cs},
\end{equation}
is the dimensionless vorticity,
\begin{equation}
\rb\equiv\frac{GM}{\cs^2}
\end{equation}
is the Bondi radius of the accreting object, and the function
$f(\oms)$ is given in terms of an integral in
\citet{krumholz05b}. For our purposes it is convenient to approximate
the integral by
\begin{equation}
\label{fomsdefn}
f(\oms) \approx
\frac{1}{1+\oms^{0.9}},
\end{equation}
which is accurate to better than $12\%$ for $\oms < 10^4$.
Equation (\ref{bhturbformula}) simply computes the harmonic mean of
the squared accretion rates, effectively minimizing between them.

We could test our \textit{ansatz} by simulating regions with a range
of vorticities and velocities, but that parameter space is relatively
large and exploring it would be time-consuming. Instead, we can test
our model directly against a turbulent region. If such a region has
volume $V$, then we predict that
\begin{equation}
\label{probpredict}
P_{\dot{M}}(<\dot{M}) = \frac{1}{V} \int
H[\dot{M}-\mdotturb(\mathbf{x})] \,dV,
\end{equation}
where $H(x)$ is the Heaviside step function, which is unity for $x>0$
and zero for $x<0$. In contrast, the model proposed by
\citet{padoan05} suggest using $\mdotbh(\mathbf{x})$ instead of
$\mdotturb(\mathbf{x})$. Note that this equation implicitly assumes
that the we can estimate the accretion rate onto a particle by looking
at the distribution of gas properties in space rather than the
distribution of gas properties in time at the particle's location,
which might seem more intuitive. However, since turbulent media are
roughly homogenous in time and space (at least over time scales
shorter than the time it takes the turbulence to decay and on length
scales smaller than the outer scale of the turbulence) the space and
time distributions should be approximately the same. We can test this
homogeneity approximation by comparing our predictions to simulations,
as we do in \S~\ref{sim}. Because we are assuming that the spatial and
temporal distribution of accretion rates are the same, for much of
what follows we will not explicitly distinguish between the two.

\section{Simulations}
\label{sim}

\subsection{Simulation Methodology}

To test our model, we simulate accretion onto point
particles in a turbulent medium. Our calculations use our
three-dimensional adaptive mesh refinment (AMR) code to solve the
Euler equations of compressible gas dynamics
\begin{eqnarray}
\frac{\partial\rho}{\partial t} + \nabla\cdot\left(\rho
\mathbf{v}\right) & = & 0 \\
\frac{\partial}{\partial t} \left(\rho \mathbf{v}\right) + \nabla
\cdot \left(\rho \mathbf{vv}\right) & = & -\nabla P -
\rho\nabla\phi \\
\frac{\partial}{\partial t}\left(\rho e\right) + \nabla \cdot
\left[\left(\rho e + P\right)\mathbf{v}\right] & = & -\rho
\mathbf{v} \cdot \nabla\phi,
\end{eqnarray}
where $\rho$ is the density, $\mathbf{v}$ is the vector
velocity, $P$ is the thermal pressure (equal to $\rho c_s^2$ since we
adopt an isothermal equation of state), $e$ is the total
non-gravitational energy per unit mass, and $\phi$ is the gravitational 
potential. The code solves these equations using a
conservative high-order Godunov scheme with an optimized approximate
Riemann solver \citep{toro97}, and its implementation is described in
detail in \citet{truelove98} and \citet{klein99}. The algorithm is
second-order accurate in both space and time for smooth flows, and it
provides robust treatment of shocks and discontinuities. We adopt a
near-isothermal equation of state, so that the ratio of specific heats
of the gas is $\gamma=1.001$.

Although the code is capable
of solving the Poisson equation for the gravitational field $\phi$ of
the gas based on the density distribution, as discussed above we
neglect the self-gravity of the gas and consider only the
gravitational potential of the accreting particles. Thus, the
potential is given by
\begin{equation}
\phi = \sum_{i=1}^{n_{\rm part}} \frac{G
M_i}{\left|\mathbf{x}-\mathbf{x}_i\right|},
\end{equation}
where $M_i$ and $\mathbf{x}_i$ are the mass and position of particle
$i$. The particles themselves are Lagrangian sink particles
implemented using the algorithm of \citet{krumholz04}. They are
capable of moving through the gas, accreting from it, and interacting
gravitationally with the gas. Although ordinarily sink particle can
also gravitationally interact with one another, in this simulation we
neglect inter-particle gravitational forces. 

The entire code operates within the AMR framework \citep{berger84,
berger89, bell94}. In AMR, one discretizes the problem domain onto a
base, coarse grid, which we call level 0. The code then dynamically
creates finer levels $L=1,2,\ldots n$ within this based on
user-specified criteria. The cells on level $L$ have cell spacings
that are a factor of $f$ smaller than those on the next coarser level, 
i.e. $\Delta x_{L} = \Delta x_{L-1}/f$, where $f$ is also
user-specified. For the runs in this paper, we use $f=2$. To take a
time step, one advances level 0 through a time $\Delta t_0$, and then
advances level 1 through $f$ steps of size $\Delta t_1 = \Delta
t_0/f$. However, for each advance on level 1, one must advance level 2
through $f$ time steps of size $\Delta t_2 = \Delta t_1/f$, and so
forth to the finest level present. The flux across a boundary between
level $L$ and level $L+1$ computed during one time step on level $L$
may not match that computed over $f$ time steps on level $L+1$. For
this reason, at the end of each set of $f$ fine advances, the code
performs a synchronization procedure at the boundaries between coarse
and fine grids to ensure conservation of mass, momentum, and energy.

\subsection{Simulation Setup}
\label{simsetup}

The initial conditions for our simulation consists of a box of gas
with a uniform density $\rho=1$ (we use non-dimensional units
throughout) and sound speed $\cs=1$. The box has periodic boundary
conditions and extends from $-1$ to $1$ in the $x$, $y$, and $z$
directions. We impose an initial turbulent velocity field by
generating a Gaussian-random grid of vectors \citep{dubinski95}. The
grid has all its power at wavenumbers $\tilde{k}=1-2$, where for a
wave of wavelength $\lambda$, $\tilde{k}\equiv 4/\lambda$. Thus,
$\tilde{k}=1$ corresponds to the largest wave that will fit in the
box. We normalize the initial velocity so that the 3-D Mach number is
$\calm_0=40$. Once the we have established the initial conditions, we
allow the turbulence to decay freely until the Mach number reaches
$\calm_0=5.0$. The decay allows the turbulence to cascade down to
small scales and reach the commonly observed $k^{-2}$ spectrum for
supersonic turbulence, which is the natural consequence of Burger's
equation. During this phase of the simulation we use a
resolution of $N=512$ cells per linear dimension with a fixed
(non-adaptive) grid. We choose 40 as the initial Mach number because
that allows us to let the turbulence decay through more than one
e-folding before we measure its properties, ensuring that the flow has
reached statistical equilibrium.

Once the turbulence has decayed to $\calm_0=5.0$, we insert sink
particles into the simulation. The particles are arrayed in a uniformly
spaced $4\times 4\times 4$ grid, and have masses $M=13/32$ (in units
where $G=1$), giving them Bondi-Hoyle radii at $\calm_0=5.0$ of
\begin{equation}
\rbh = \frac{M}{1+\calm_0^2} = \frac{1}{64}.
\end{equation}
Since the particles are separated from their nearest neighbors by a
distance of $1/2$, they are separated by $32\rbh$ and
should accrete completely independently of one another. When we
introduce the particles, we refine the regions around them to
guarantee that at every point in space $r/\Delta x > 16$, where $r$ is
the distance to the nearest particle and $\Delta x$ is the grid
spacing of the finest grid covering that point. We continue this
refinement to a maximum resolution of $N=8192$, giving a grid spacing
of $\Delta x = 1/4096 = \rbh/64$ on the finest level. We discuss
issues of resolution and convergence in more detail in
\S~\ref{resconverge}.

\subsection{Simulation Results}
\label{bhturbsimresult}

\begin{figure}
\epsfig{file=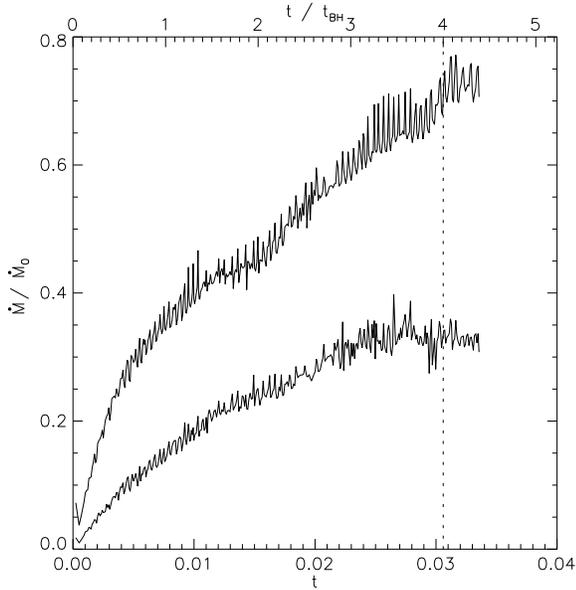}
\caption{\label{mdottime}
Mean (\textit{upper line}) and median (\textit{lower line}) accretion
rate versus time. The dotted vertical line shows the time when the
accretion rate has reached equilibrium.
}
\end{figure}

After we have inserted the sink particles, we allow a simulation to
evolve until both the median and mean accretion rates (mean and median
over particles, not over time) onto the
particles becomes roughly constant, as shown in Figure
\ref{mdottime}. We estimate by eye that the accretion rate has
stabilized at time $t>4\tbh$, where
\begin{equation}
\tbh\equiv \frac{\rbh}{\calm_0 \cs} = 0.00313.
\end{equation}
The accretion rate in the plot is normlized to $\mdotzero=0.0166$, in
our dimensionless units.

\begin{figure}
\epsfig{file=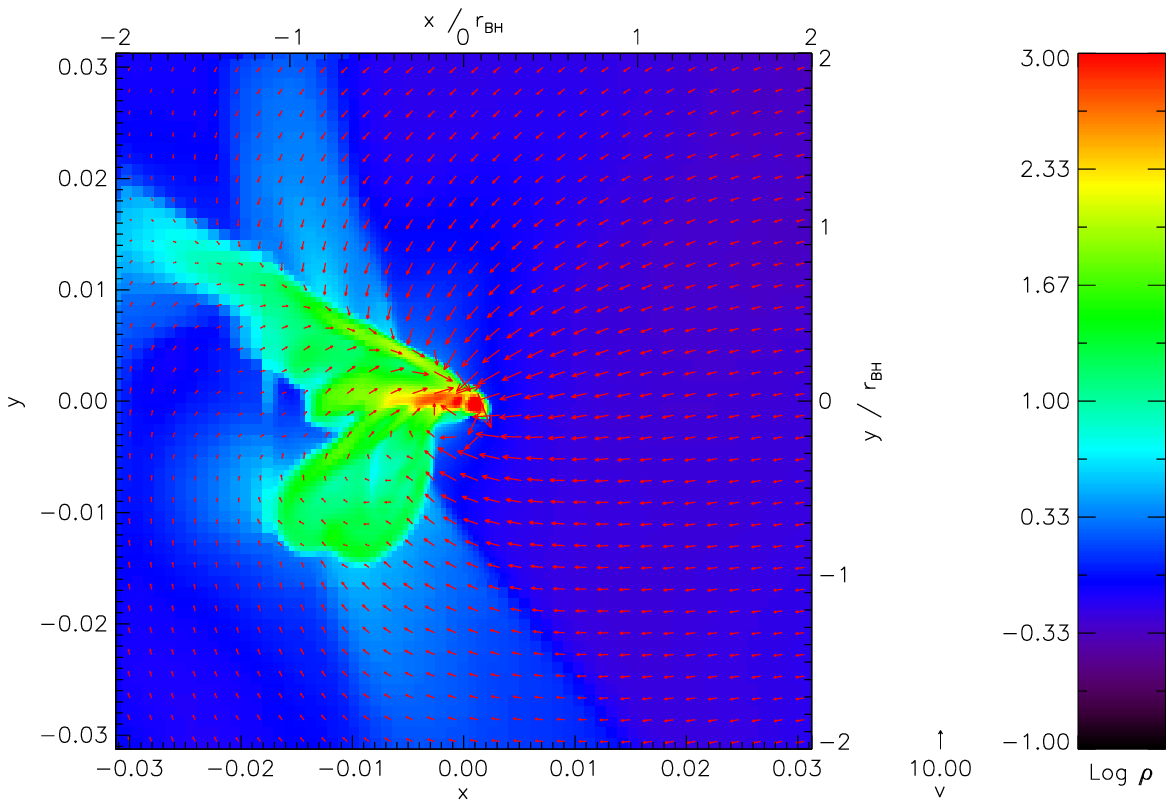, scale=0.7}
\caption{\label{slicebh}
Density (\textit{color}) and velocity (\textit{arrows}) around an
accreting particle. The origin is centered on the accreting
particle, and the axes show position in code units and in units of
$\rbh$.
}
\end{figure}

\begin{figure}
\epsfig{file=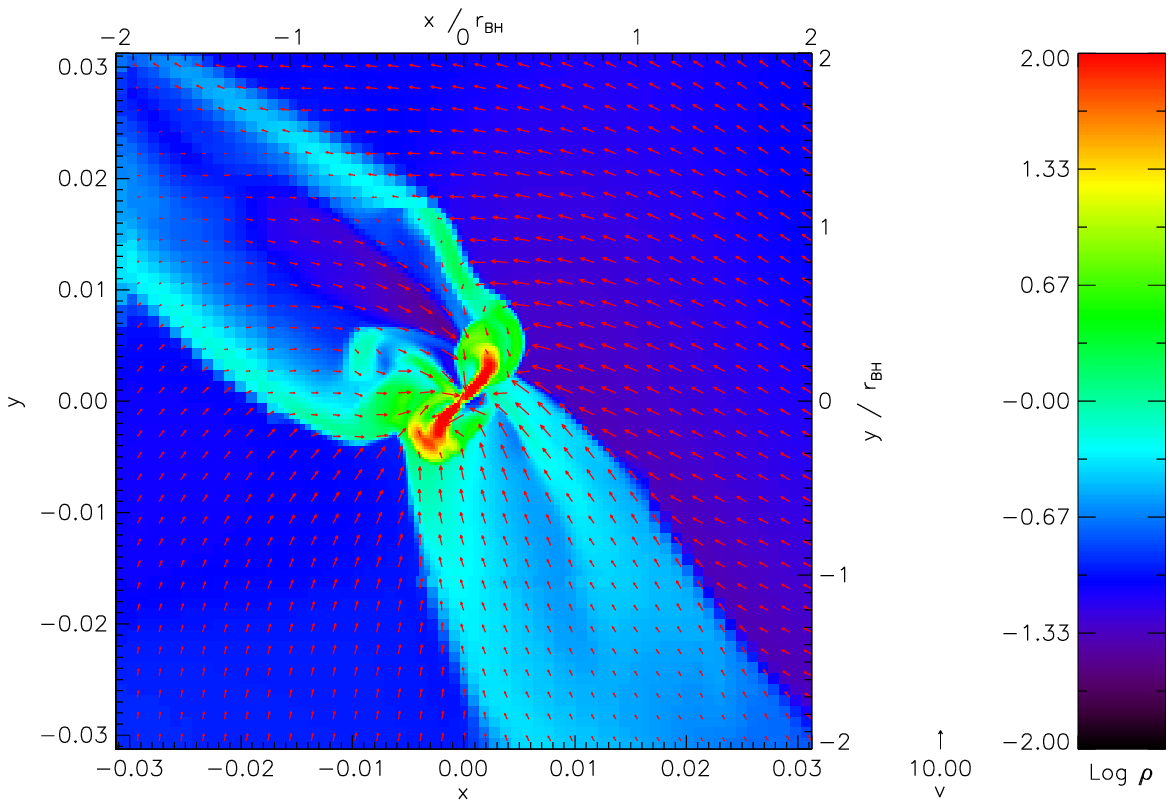, scale=0.7}
\caption{\label{slicevort}
Density (\textit{color}) and velocity (\textit{arrows}) around an
accreting particle. The origin is centered on the accreting
particle, and the axes show position in code units and in units of
$\rbh$.
}
\end{figure}

As our simple model predicts, after the accretion rate has reached
equilibrium the flow pattern around some of the
particles is similar to that of Bondi-Hoyle accretion with no
vorticity, and the flow pattern around others is closer to the case of
high vorticity. Figure \ref{slicebh} shows an example of the
former. The velocity of the gas on the right side of the plot is
relatively large and relatively uniform, indicating there is little
vorticity. As the gas passes the sink particle, the sink particle's
gravity causes streamlines to converge and a shock to form. This leads
to a clear Mach cone, as is normally seen in Bondi-Hoyle accretion
\citep[e.g.][Figure 3]{krumholz04}. In contrast, Figure
\ref{slicevort} shows a vorticity-dominated accretion flow. There is
no Mach cone, but there is a dense, rotationally-supported torus of
gas around the accreting particle. This is similar to the flow
patterns that occur for accretion with no net relative velocity
between the gas and the particle, just vorticity \citep[e.g.][Figure
5]{krumholz05b}.

\begin{figure}
\epsfig{file=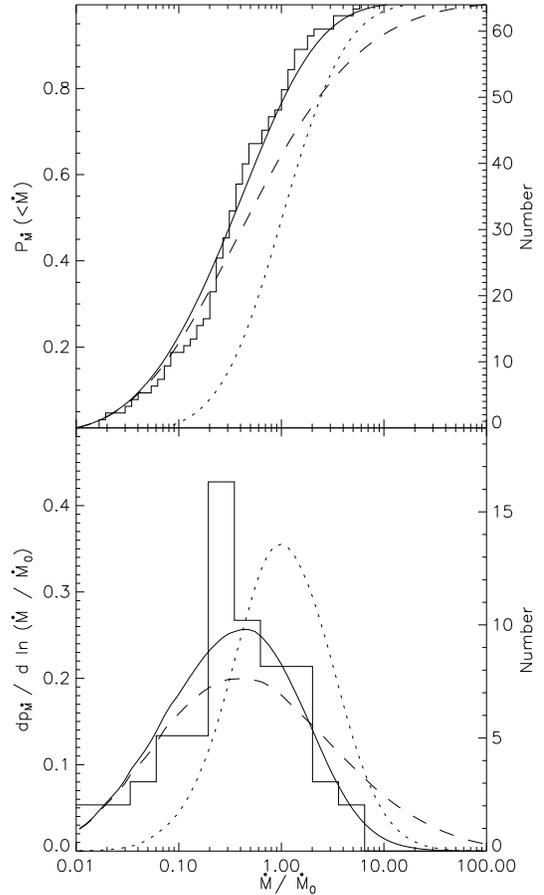}
\caption{\label{pdfplot}
Cumulative (\textit{upper panel}) and differential (\textit{lower
panel}) accretion rate PDFs
vs. $\dot{M}/\dot{M}_0$. The curves shown are the measured value from
the simulation (\textit{histogram}) and the theoretical predictions
using $\mdotturb$ (\textit{solid line}), $\mdotbh$ (\textit{dashed
line}), and $\mdotomega$ (\textit{dotted line}).
}
\end{figure}

We next compare the accretion rate predicted by our model to what we
measure in the simulations. To measure the accretion rate in the
simulations, we wait until $t>4\tbh$, and compute the time-averaged
accretion rate onto each particle thereafter. We then construct the
cumulative
distribution function of the accretion rates onto the particles, which
we wish to compare with $P_{\dot{M}}(<\dot{M})$ as predicted by our
models. For the model, we examine the simulation cube at the time when
we insert the sink particles. We then compute $\mdotbh$ (equation
\ref{bhformula1} with $\lambda = \exp(1.5)/4$),
$\mdotomega$ (equation \ref{mdotomegaformula}) and $\mdotturb$
(equation \ref{bhturbformula}) in every cell of the simulation. To
measure vorticity, which is not a primitive of our simulation, we take
two-sided differences over one cell at every point. From
the grid of accretion rates, we compute $P_{\dot{M}}(<\dot{M})$, using
$\mdotbh$, $\mdotomega$, and $\mdotturb$ in equation
(\ref{probpredict}). We compare the predictions to the measured
accretion rates in Figure \ref{pdfplot}.

Note that, since $\tbh\ll 1$,
the state of the simulation cube on the large scale does not change
much over the time after we insert the sink particles. For example, if
we continue to let the box evolve without adding sink particles, after
$4\tbh$, the Mach number only decreases from $5.0$ to $4.7$. Thus,
there is no problem in applying our simple model at the time just
before we insert the sink particles. Also note that, at the time we
measure vorticity, we have not yet inserted the sink particles and so
there are no higher level AMR grids. When these grids first appear,
the velocity is linearly interpolated on a cell-by-cell basis, so the
vorticity in a refined cell is equal to that of its parent. However,
as the simulation evolves, baroclinic instability generates vorticity
structure on more refined levels just as it does on the base
level. By the time our accretion rates stabilize, the vorticity
structure on the more refined levels has had several crossing-times to
reach equilibrium.

As Figure \ref{pdfplot} shows, our \textit{ansatz} function
$\mdotturb$ gives excellent agreement with the simulation results. In
contrast, using either $\mdotbh$ or $\mdotomega$ in place of
$\mdotturb$ gives an extremely poor fit. We quantify this result by
comparing the theoretical models to the simulations using a
Kolmogorov-Smirnov (KS) test \citep{press92}. The KS statistic for our
model is $0.81$. For the
\citet{padoan05} proposal of using $\mdotbh$ where we use $\mdotturb$,
it is $0.012$, and using $\mdotomega$ instead of
$\mdotturb$ gives $2.7\times 10^{-10}$. This indicates that our model is
highly consistent with the simulation data, while the two other
candidates we have checked are essentially ruled out. The difference
is large enough to have substantial astrophysical significance: the
accretion rate averaged over particles that we measure in our
simulations is $\langle\dot{M}\rangle/\dot{M}_0 = 0.73$, and our
theory predicts a mean of $\langle\dot{M}\rangle/\dot{M}_0 =
0.90$. (The difference is larger than one might expect because our 64
particles do not sample the tail of the distribution very well.) In
contrast, the \citet{padoan05} theory predicts
$\langle\dot{M}\rangle/\dot{M}_0 = 4.3$, an ovestimate of
nearly an order of magnitude.

\begin{figure}
\epsfig{file=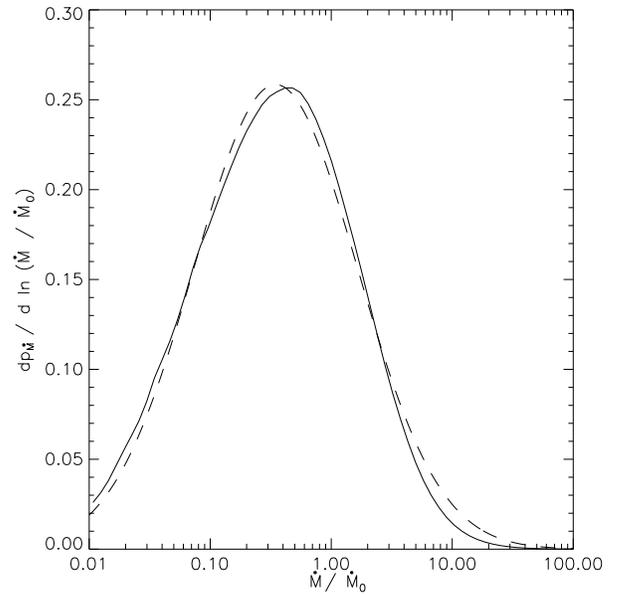}
\caption{\label{pdffit}
Numerical (\textit{solid line}) and best-fit theoretical model
(\textit{dashed line})
accretion rate probability distributions vs. $\dot{M}/\dot{M}_0$. 
}
\end{figure}

Our theoretical model is well-described by a fit to a lognormal
PDF. We find a best fit of
\begin{equation}
\frac{dp_{\dot{M}}}{d\ln \dot{M}} \approx
0.26 \, \exp
\left\{-\frac{\left[\ln\left(\dot{M}/\dot{M}_0\right)-1.1\right]^2}
{2(1.6)^2}\right\}.
\end{equation}
Figure \ref{pdffit} shows our fit in comparison to the simulation
result.

A final question to consider is whether we might be underestimating
the accretion rate because we only count mass that falls onto the
particle, not mass that accumulates in the rotationally-supported
torus or other bound structures around the particle. One might worry
that these structures are acting as mass reservoirs, and that the
accretion rate might rise at some point in the future as mass accretes 
from them on a viscous time scale. A priori this seems unlikely:
\citet{krumholz05b} simulate accretion with vorticity for 200 Bondi
times and find that the rotationally-supported toruses do not act as 
reservoirs that raise the accretion rate at later times. Instead, the
pattern is that the accretion rate starts high and gradually
declines to its equilibrium value, while the toruses reach a steady
mass, so that the ratio of torus mass to accreted mass declines
steadily with time. Real accreting objects are likely to be in this
limit where the torus mass is negligible compared to the total
accreted mass. For example, the Bondi-Hoyle time for a protostar
accreting in from a turbulent molecular clump is, as we show in
\S~\ref{implications}, only $\sim 10$ yr, vastly shorter than the time
for which the protostar accretes.

We unfortunately cannot run our simulations for hundreds of
Bondi-Hoyle times due to limitations of computational resources, and
because over such long time scales the particles would likely
interfere with one another. Thus, we cannot do a direct test as in
\citet{krumholz05b}. However, to eliminate the possibility that
we are missing substantial accretion by neglecting the gas in bound
structures, we compute all the mass within a distance
$\rbh$ of each sink particle that has a potential energy larger than
its kinetic energy, i.e. that is bound to that sink particle. We do
this at $4\tbh$, when the particle accretion rate has reached
equilibrium, and at the final time in the simulation, roughly
$4.4\,\tbh$. We then 
compute the difference between the bound mass at $4\tbh$ and at the
final time. The median change in the mass of bound structures, divided
by the median change in mass accreted onto particles over the same
period, is 0.27. Thus, even if we were to assume that bound structures
were reservoirs of mass that grow with time, the extra accretion onto
them would only represent a $\sim 30\%$ correction to the accretion
rate we have computed. If we add the accretion rate onto bound
structures to the accretion rate onto particles, and compare that to
our theoretical model using a KS test, we find a statistic of
0.15. While this is not quite as good as our value obtained using just
the particles, this is still a reasonably good fit.

Note, however, that including the mass in bound structures only
makes sense if they are truly accreting. Otherwise, adding their mass
to the mass accreted by the particles is simply adding noise. It seems
likely that this is the case, because the bound mass fluctuates
strongly, and can be quite small compared to the accreted mass: for 16
of our 64 particles, the bound mass is less than 10\% of the accreted
mass at some point during the simulation, for 32 of the particles it
is less than 25\% of the accreted mass, and for 58 of them it is less
than the accreted mass. Since the amount of mass that remains in the
toruses over long times is small compared to the accreted mass, the
toruses cannot be acting as mass reservoirs. Instead, the mass within
them is transient, either on its way to being accreted or on its way
out of the Bondi-Hoyle radius, not accumulating over a long term. Our
estimates of the accretion rate using just the mass that falls onto
the particles are more accurate.

\section{The Accretion Rate in Turbulent Flows}
\label{extension}

Thus far we have shown that our model very accurately predicts the
distribution of accretion rates in terms of the density and velocity
field of the gas. We can use this result to derive an estimate for the 
accretion rate in a general turbulent flow in terms of a few simple
parameters of the turbulent region that can usually be determined from 
observations. We consider a turbulent region of characteristic size
$\ell$, mean density $\overline{\rho}$, sound speed $\cs$, and
three-dimensional Mach number $\calm_0$, accreting onto an object of
mass $M$. We derive a theoretical estimate for the mean and median
accretion rate based on the probability distributions of density,
velocity, and vorticity in a turbulent medium in
\S~\ref{extensiontheory}, and we compare it to simulations in
\S~\ref{extensionsim}. We then discuss the issue of differences
between mean and median accretion rates in \S~\ref{medianvsmean}. In
this Section we limit ourselves to discussing accretors randomly
sampling turbulent regions. We discuss how to apply this results in
real astrophysical situations in \S~\ref{limitations} and
\S~\ref{implications}.

\subsection{Theoretical Calculation}
\label{extensiontheory}

Although density, velocity, and vorticity in a turbulent medium are
certainly correlated to some extent, that correlation is relatively
weak. We can derive reasonable estimates for the functional dependence 
of the accretion rate on the properties of a turbulent region simply
by neglecting correlations and assuming that density, velocity, and
vorticity are selected independently from the appropriate PDF at every 
point.

\begin{figure}
\epsfig{file=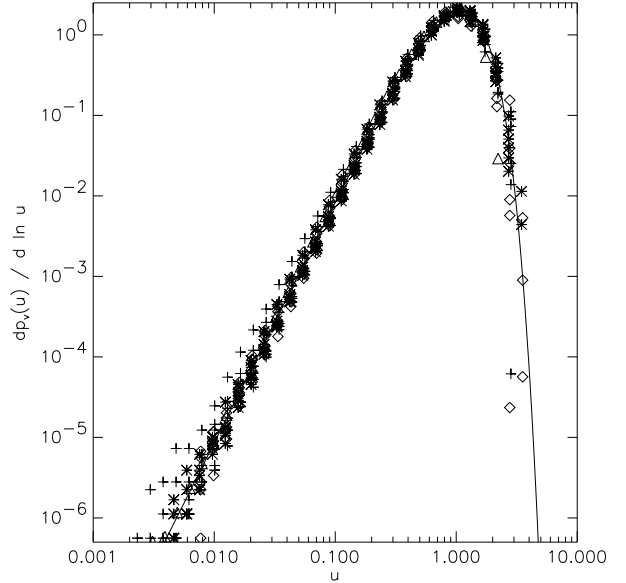}
\caption{\label{velpdfplot}
Measured values and fit for $dp_v(u)/d\ln u$. We show measurements
from our 6 $N=256$ runs at $\calm_0=3$ (\textit{diamonds}),
$\calm_0=5$ (\textit{asterisks}), and $\calm_0=10$ (\textit{crosses}),
from our $N=512$ run at $\calm_0=5$ (\textit{triangles}), and our best 
fit (\textit{line}).
}
\end{figure}

Numerous authors have studied the PDF of densities in a turbulent
medium \citep{vazquezsemadeni94, padoan97, scalo98, passot98,
nordlund99, ostriker99, padoan02}. \citet{padoan02} find that the PDF is
well-fit by the functional form
\begin{equation}
\label{rhopdf}
\frac{dp_{\rho} (x)}{d\ln x} = 
\frac{1}{\sqrt{2\pi \sr^2}} \exp\left[-\frac{\left(\ln x -
\overline{\ln x}\right)^2}{2\sr^2}\right]
\end{equation}
where $x=\rho/\overline{\rho}$ is the density normalized to the mean
density. The mean of the log of density (and also the log of the
median density) is
\begin{equation}
\overline{\ln x} = -\frac{\sr^2}{2}.
\end{equation}
and the dispersion of the PDF is approximately
\begin{equation}
\label{sreqn}
\sr \approx \left[\ln\left(1+\frac{\calm_0^2}{4}\right)\right]^{1/2},
\end{equation}
The PDF of velocities is considerably less well-studied in the
literature. Measuring from the simulations we describe in
\S~\ref{extensionsim}, we find that the PDF is well-fit by a power law 
with an exponential cutoff,
\begin{equation}
\frac{dp_v(u)}{d\ln u} \propto u^3 \exp\left(-1.5\, u^{1.7}\right),
\end{equation}
where $u=v/(\calm_0\cs)$ is the velocity normalized to the velocity
dispersion of the region. Figure \ref{velpdfplot} shows our
measurements from the simulations and our fit. The median velocity for 
this PDF is $1.2 \,\calm_0\cs$, but the factor of 1.2 is
relatively uncertain because it depends strongly on the form of the
exponential cutoff, which is poorly constrained in our
fit. We therefore simply take the median velocity to be
$\mbox{median}(v)\equiv \phi_{\rm v} \calm_0 \cs$, where $\phi_{\rm
v}$ is a constant near unity.

\begin{figure}
\epsfig{file=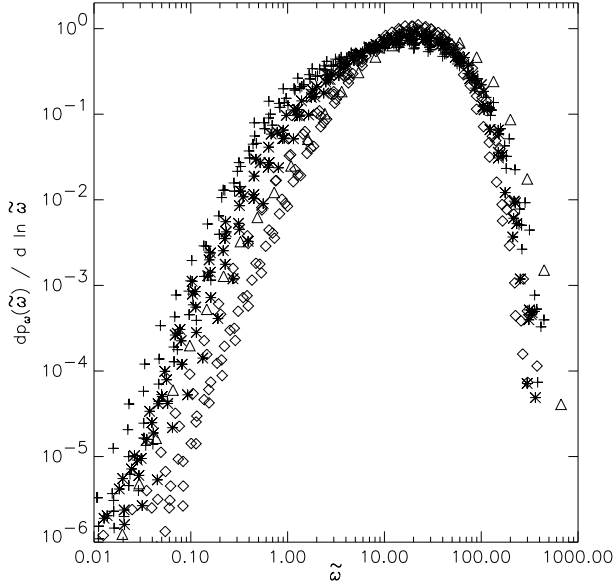}
\caption{\label{vortpdfplot}
Measured values and estimated median for
$dp_{\omega}(\tilde{\omega})/d\ln \tilde{\omega}$. We show measurements
from our 6 $N=256$ runs at $\calm_0=3$ (\textit{diamonds}),
$\calm_0=5$ (\textit{asterisks}), and $\calm_0=10$ (\textit{crosses}),
and from our $N=512$ run at $\calm_0=5$ (\textit{triangles}).
}
\end{figure}

The vorticity PDF is considerably more complex, as shown in Figure
\ref{vortpdfplot}. In making the Figure we have introduced the
normalized vorticity
\begin{equation}
\tilde{\omega}=\omega \frac{\ell}{\calm_0\cs},
\end{equation}
which is related to the vorticity parameter for accretion onto an
object by $\oms=\tilde{\omega}(\calm_0 \rb/\ell)$.
The vorticity PDF shows substantial variation from run to run, the
shape does not appear to be independent of Mach number, and the shapes
are not easily fit by a simple functional form. However, we can still
obtain useful information by noting that the PDF peaks at 
roughly the same location independent of Mach number, and the median
normalized vorticity is roughly constant. We estimate it as
$\mbox{median}(\tilde{\omega}) \equiv 10\,\phi_{\omega}$, where
$\phi_{\omega}$ is a constant of order unity. Note that the fact that
this peak is roughly independent of Mach number means that the median
vorticity in a supersonically turbulent region is completely specified
by its velocity dispersion and physical size.

From these PDFs, we can estimate the median accretion rate. Since 
we only expect the fit to be approximate, we simplify the results by
setting $\lambda=1$ in equation (\ref{bhformula1}) and by
droping terms of order $\calm_0^{-2}$ or smaller when added to terms of
order unity. With these approximations, and neglecting correlations
between density, velocity, and vorticity so that we can treat the
median operator as linear, we predict that the median accretion rate,
normalized to our simple Bondi-Hoyle estimate, will be
\begin{eqnarray}
\phimed & \equiv & \mbox{median}(\dot{M})/\dot{M}_0 \\
& \approx &
\mbox{median}\left[\left(\dot{M}_{\rm
BH}^{-2}+\dot{M}_{\omega}^{-2}\right)^{-1/2}\right] / \dot{M}_0 \\
& \approx & \frac{\phi_{\rm v}^{-3}}{\sqrt{1+\calm_0^2/4}}
\left[1 + \frac{10}{\phi_{\rm v}^6} \left(
\frac{10\,\phi_{\omega} \rb}{\ell
\calm_0^{2.3}}\right)^{1.8}\right]^{-0.5}.
\label{phimedfit}
\end{eqnarray}
The functional form agrees with what one would intuitively
expect. When $\rb/\ell$ is sufficiently small, vorticity has no effect 
on the accretion rate and all vorticity-related terms
disappear. Instead, one is left with a median accretion rate that is
close to the simple Bondi-Hoyle estimate $\dot{M}_0$. It is reduced
relative to this, however, by the factor $(1+\calm_0^2/4)^{-1/2}$,
which is the ratio of the median to the mean density. When 
$\rb/\ell$ is large enough that the term in square brackets is much
greater than unity, the median accretion rate is controlled entirely
by vorticity. Since the vorticity accretion rate depends approximately 
on $\oms^{-0.9}$, and $\oms\propto \phi_{\omega} \rb/\ell$, we expect
the accretion rate scale as $(\phi_{\omega} \rb/\ell)^{-0.9}$, which
is exactly what our estimate gives. The crossover between the two
regimes occurs roughly when $\dot{M}_{\rm BH}=\dot{M}_{\omega}$, which
occurs when the second term in the square brackets is about unity.

We can perform a similar procedure for the mean, although here we are
considerably hampered by our uncertainties about the exact functional
form of the vorticity and its dependence on Mach number. This
uncertainty is amplified by the fact that the low vorticity tail of
the PDF contributes significantly to the mean accretion rate. 
For this reason, we concentrate on estimating the mean accretion rate
in the case where vorticity is irrelevant, and in
\S~\ref{extensionsim} we perform a purely empirical fit in the
case where vorticity is important. For the case where vorticity is
irrelevant, we compute the mean accretion rate by integrating over the
density and velocity PDFs, again assuming that they are independent
and setting $\lambda=1$ in (\ref{bhformula1}). This gives
\begin{eqnarray}
\phimean & \equiv &
\left\langle \dot{M}\right\rangle /\dot{M}_0 \\
\label{bhavg}
& \approx & \int_0^{\infty} 
\frac{1}{\left(\calm_0^{-2}+u^2\right)^{3/2}}
\frac{dp_v(u)}{du} \, du.
\end{eqnarray}
Although we can evaluate this expression numerically for
our best-fit PDF, it is more convenient to approximate
the integral in the case $\calm_0\gg 1$ by treating the
exponential cutoff in $dp_v(u)/du$ as a sharp truncation that gives
unity at $u<\phi_{\rm u}$ and zero at $u>\phi_{\rm u}$, with
$\phi_{\rm u}\approx 1$. The factor $\phi_{\rm u}$ represents our
uncertainty in the exact shape of the exponential cutoff on the
velocity PDF. With these approximations, we find
\begin{equation}
\phimean
\approx \frac{3}{\phi_{\rm u}^3} \left[\ln(2\phi_{\rm u}
\calm_0) -1\right], \qquad \left(\frac{\rb}{\ell}\rightarrow 0\right).
\end{equation}
For $\phi_{\rm u}=1$, this expression agrees with (\ref{bhavg}) to
better than $10\%$ for all Mach numbers between 2 and 200.

\subsection{Comparison to Simulations}
\label{extensionsim}

We now compare our theoretically predicted mean and median accretion
rates to simulations. We run a series of
simulations of periodic boxes, using the same methodology as described
in \S~\ref{sim}, and compute the mean and median values of the
accretion rate using our simple model. To get a sense of the range of
variation in accretion rates, we run six simulations at a
resolution of $N=256$, each using a different random realization of
the initial velocity field. We examine each run at the time when the
Mach number has decayed to $\calm_0=10$, $\calm_0=5$, and finally
$\calm_0=3$. To compute the accretion rate from our simple model, we
must also specify $\rb/\ell$, the size of the particle's Bondi radius
relative to the size of the turbulent region, since the vorticity
parameter $\oms$ is proportional to $\rb$. We compute $\phimean$
and $\phimed$ for each realization, at each Mach number, and a range
of values of $\rb/\ell$. Since our theory will probably fail for
objects with Bondi-Hoyle radii comparable to the size of the entire
turbulent region (see \S~\ref{limitations}), we limit ourselves to
$\rbh/\ell \ltsim 0.1$, or $\rb/\ell\ltsim 10$ at $\calm_0=10$. 

\begin{figure}
\epsfig{file=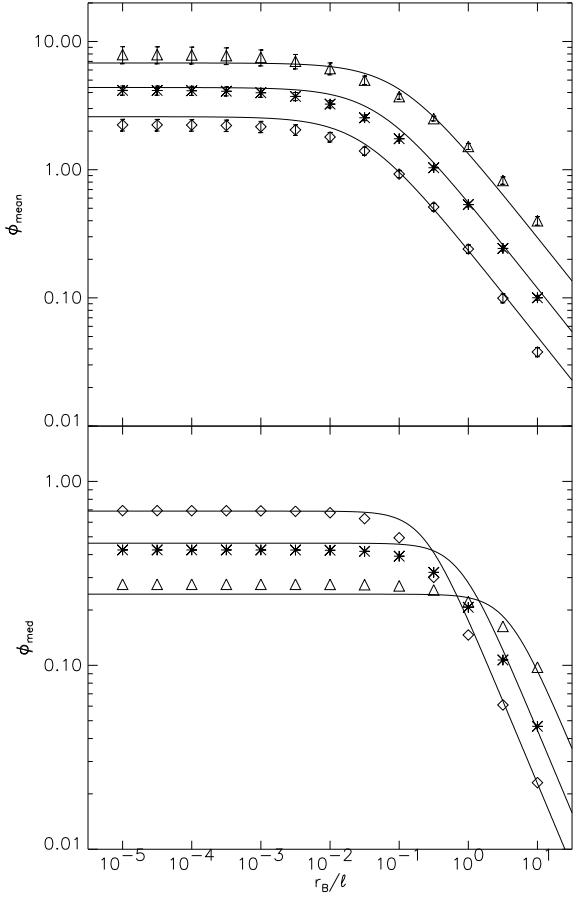}
\caption{\label{phifit}
Measured and fit values of $\phimean$ (\textit{upper panel}) and
$\phimed$ (\textit{lower panel}) vs. $\rb/\ell$. The plots show runs
Mach 3 (\textit{diamonds}), Mach 5 (\textit{asterisks}), and Mach 10
(\textit{triangles}), and our fits for these Mach numbers
(\textit{solid lines}). The error bars in the upper panel show the
standard deviations in $\phimean$. We do not show the error bars in
$\phimed$ in the lower panel to avoid cluttering the plot.
}
\end{figure}

We report the mean and standard deviation of the six runs, for
selected values of $\rb/\ell$, in Table \ref{phitable}, and plot the
results in Figure \ref{phifit}. The Figure also shows fits based on
our theoretical predictions for $\phimed$ and $\phimean$. For
$\phimed$, we find a good fit with equation (\ref{phimedfit}) for
$\phi_{\rm v}=0.93$ and $\phi_{\omega} = 1.25$. For $\phimean$, we
only have a theoretical prediction for the low $\rb/\ell$ case, so we
fit to a function of the form $\phimean = (\langle\dot{M}_{\rm
BH}\rangle/\dot{M}_0) [1  + f(\rb/\ell, \calm_0)]^q$, where
$f(\rb/\ell, \calm_0)$ is a power-law function of $\rb/\ell$ and
$\calm_0$. We find that the function
\begin{equation}
\label{phimeanfit}
\phimean = \frac{3}{\phi_{\rm u}^3} \left[\ln(2\phi_{\rm
u}\calm_0)-1\right]
\left(1+100\frac{\rb}{\ell \calm_0}\right)^{-0.68}
\end{equation}
with $\phi_{\rm u}=0.95$ fits the data reasonably well. Overall, we
find that our predicted functional forms provide a good fit to
the simulation data and capture the essential physics of the accretion 
process.

\begin{deluxetable}{cccc}
\tablecaption{Selected computed accretion rates.\label{phitable}}
\tablewidth{0pt}
\tablehead{
\colhead{$\calm_0$} &
\colhead{$\log (\rb/\ell)$} &
\colhead{$\phimean$} 
&
\colhead{$\phimed$} 
}
\startdata
3   & $-5.0$ &  $ 2.2 \pm 0.23$ &  $ 0.69 \pm 0.14$ \\ 
3   & $-3.0$ &  $ 2.2 \pm 0.22$ &  $ 0.69 \pm 0.14$ \\ 
3   & $-1.0$ &  $ 0.93 \pm 0.059$ &  $ 0.49 \pm 0.092$ \\ 
3   & $ 1.0$ &  $ 0.038 \pm 0.003$ &  $ 0.023 \pm 0.003$ \\ 
5   & $-5.0$ &  $ 4.2 \pm 0.34$ &  $ 0.42 \pm 0.12$ \\ 
5   & $-3.0$ &  $ 4.0 \pm 0.31$ &  $ 0.42 \pm 0.12$ \\ 
5   & $-1.0$ &  $ 1.7 \pm 0.072$ &  $ 0.39 \pm 0.11$ \\ 
5   & $ 1.0$ &  $ 0.10 \pm 0.004$ &  $ 0.047 \pm 0.008$ \\ 
10  & $-5.0$ &  $ 7.9 \pm 1.2$ &  $ 0.28 \pm 0.074$ \\ 
10  & $-3.0$ &  $ 7.5 \pm 1.1$ &  $ 0.28 \pm 0.074$ \\ 
10  & $-1.0$ &  $ 3.7 \pm 0.18$ &  $ 0.27 \pm 0.071$ \\ 
10  & $ 1.0$ &  $ 0.40 \pm 0.031$ &  $ 0.098 \pm 0.016$ \\ 
\enddata
\tablecomments{
Cols. (3-4): Results are reported as 
$(\mbox{mean}) \pm (\mbox{standard deviation})$.
}
\end{deluxetable}

Since the distribution of accretion rates is approximately lognormal,
we can turn our fits for $\phimean$ and $\phimed$ into fits for the
entire PDF of accretion rates. A lognormal distribution
is completely characterized by two parameters, the center of the
distribution and its width, and we can solve for these in terms of
$\phimean$ and $\phimed$. Doing so, we find that the probability
distribution of accretion rates in a turbulent medium is
\begin{eqnarray}
\lefteqn{\frac{dp_{\dot{M}}}{d\ln (\dot{M}/\dot{M}_0)} =
\frac{1}{\sqrt{2\pi\smdot^2}} \cdot}
\nonumber \\
& & \exp\left\{
-\frac{\left[\ln (\dot{M}/\dot{M}_0) -
\overline{\ln(\dot{M}/\dot{M}_0)}\right]^2} {2\smdot^2}
\right\},
\end{eqnarray}
where 
\begin{equation}
\overline{\ln(\dot{M}/\dot{M}_0)} = \ln \phimed
\end{equation}
and
\begin{equation}
\smdot = \sqrt{2 \ln\frac{\phimean}{\phimed}}.
\end{equation}

\subsection{Median versus Mean Accretion Rates}
\label{medianvsmean}

Since the volumetric median and mean accretion rates we have found can
be quite different, we would like to determine whether one should use
a median or a mean accretion rate in attempting to follow the
evolution of an individual object. The mean and median are different
because rare, high accretion regions contribute significantly to the
overall accretion rate, even though it is very unlikely for a randomly 
placed accretor to be in one. When an object begins accreting in a
turbulent medium, its accretion rate is most likely to be near the
median. Over time, however, it will sample more and more of the
turbulent medium, and its time-averaged accretion rate should approach 
the volume average for the region. We wish to determine how long this
will take, since it is possible that it may be longer than the typical 
time that the object will spend in the turbulent region. This will
tell us if our volumetric mean and median are truly good
approximations for the mean and median accretion rate in time,
measured for a single object rather than an ensemble.

\begin{figure}
\epsfig{file=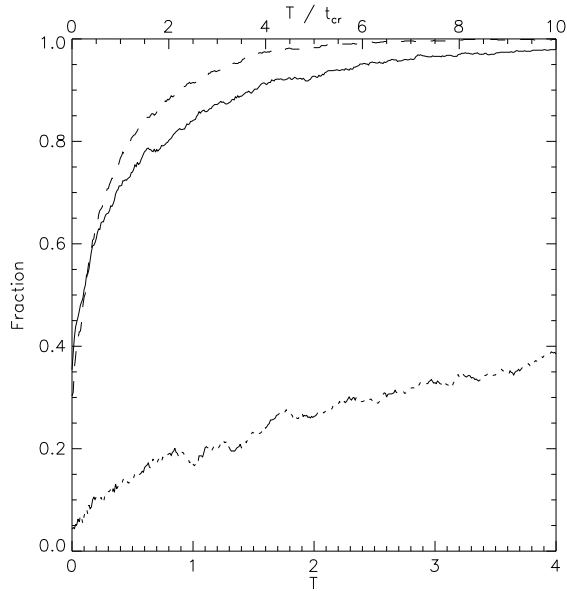}
\caption{\label{pathfrac}
Fraction of particles vs. time. The curves show the fraction
closer to the mean than the median (\textit{solid line}), within a
factor of $2.0$ of the mean accretion rate (\textit{dashed line}), and 
within a factor of $1.1$ of the mean accretion rate
(\textit{dot-dashed line}).
}
\end{figure}

To answer to the question of how long an object must accrete before
reaching the volume mean accretion rate, we consider the related
question of how far an object must move through a turbulent volume of
gas before reaching the mean accretion rate. We answer this using the
density and velocity field of our $N=512$ run at the point when
$\calm_0=5.0$, just before we insert the sink particles. The rate at
which an accreting particle moves through the volume, accreting from
new regions, should be roughly equal to the velocity dispersion of the 
gas. Thus, we place a particle at a random point within our turbulent
box, moving in a random direction with velocity $\mathbf{v}$ such that 
$\left|\mathbf{v}\right|=5.0$, while the density and velocity of the gas
remain fixed. We then compute the time-averaged accretion rate along the
particle's trajectory,
\begin{equation}
\left\langle\dot{M}\right\rangle = \frac{1}{T}\int_0^T
\mdotturb(\mathbf{x}+\mathbf{v}t) \,dt,
\end{equation}
as a function of time $T$. We repeat this procedure 1000 times. Figure 
\ref{pathfrac} shows the fraction of particles for which the
time-averaged accretion rate is closer to the volume-averaged mean
than to the median as a function of time. (We use closer in a
logarithmic sense, i.e. $\dot{m} >
\sqrt{\phimean\phimed}$.) The Figure also shows the fraction of
particles whose time-averaged accretion rates are within a factor of 1.1
and within a factor of 2.0 of the mean accretion rate in the volume. A 
majority of the particles have time-averaged accretion rates closer to
the mean than the median after only $0.25$ crossing times, where
$t_{\rm cr} \equiv \ell/(\calm_0\cs) = 0.4$, but $95\%$
do not become closer to the mean than to the median until more than
$6$ crossing times. Similarly, more than half of the particles are
within a factor of $2.0$ of the mean accretion rate after only $0.28$
crossing times, but even after 10 crossing times fewer than half have
time-averaged accretion rates within $10\%$ of the mean. Our
conclusion is that one should use the mean rather than the median
accretion rate to follow the evolution of any object that accretes for
more than a quarter of a crossing time, but that one should not expect 
the time-averaged accretion rate to be closer than tens of percent to
that value unless the particle accretes for a very long time.

\section{Discussion}
\label{bhtdiscussion}

\subsection{Resolution and Convergence}
\label{resconverge}

It is critical to establish that our simulations have sufficient
resolution for us to believe our results. We first address whether we
have enough resolution to compute the correct accretion rates onto our
particles. Our simulation resolves the Bondi-Hoyle radius of the
particles by 64 cells, and their Bondi radius by almost 1700
cells. \citet{krumholz04} found using the same code that we use here
that in simulations of Bondi-Hoyle accretion a resolution of $\Delta x
= \rbh/50$ is more than adequate to reproduce the value found by previous
theoretical and numerical work. Our resolution meets this requirement, 
although it is possible that for regions of the flow where the
velocity is significantly higher than the mean we might not have
sufficient resolution. At most this will affect the low accretion rate
tail of the distribution, and should therefore not substantially bias
our comparison between the accretion rate onto particles and our
theoretical model. Similarly, \citet{krumholz05b} found that $\Delta x
= \rb/160$ was sufficient to correctly model accretion with vorticity
parameters up to $\oms = 30$. The typical vorticity parameter we find
in our simulation is somewhat greater than this, but by less than an
order of magnitude. In contrast, our resolution is $\Delta x \approx
\rb/1700$, more than an order of magnitude greater than that
\citet{krumholz05b} found sufficient.

A related question is whether we have enough space between our
particles for them to accrete independently of one another. A priori
we do not expect the particles to interfere with one another, since
they are separated by $32\rbh$, much larger than any particle's range
of influence. There could potentially be interference between
particles if one particle were upstream of another and significantly
altered the flow of gas to its downstream neighbor. However, given the
random alignment ot the turbulent flow, having one particle
immediately upstream of another is extremeley unlikely. Even if one
were upstream of another, we run the simulation for a time $\ll 32
\tbh$ once we insert the particles, so there is no time for depletion
to propogate from one particle to another. Finally, we can find
evidence against the possibility of interference
between the particles by examining the accretion rates we actually
measure. For Bondi-Hoyle accretion, the accretion rate depends on the
accretion radius as $\dot{M}\propto r_{\rm acc}^{3/2}$, and a particle 
accreting at rate $\dot{M}_0$ has an accretion radius $r_{\rm
acc}\approx\rbh$. We would expect the particles to begin interfering with
one another when the accretion radius is comparable to the half the
inter-particle separation, or $16\rbh$. This corresponds to an
accretion rate of $64\dot{M}_0$. If interference between particles
were a problem, we would expect to see the distribution of accretion
rates truncated around $64\dot{M}_0$. However, Figure \ref{pdfplot}
shows that there  are no particles accreting at anywhere near this
rate, and that even our theoretical distribution predicts essentially
no particles accreting at such high rates. We therefore conclude that
none of our particles have accretion radii large enough for them to
interfere with one another.

\begin{figure}
\epsfig{file=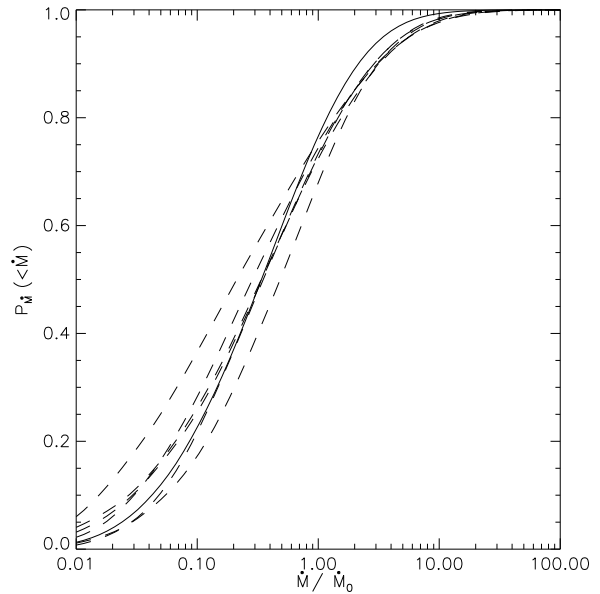}
\caption{\label{convpdf}
PDF of accretion rates for the $N=256$ runs (\textit{dashed lines})
and the $N=512$ run (\textit{solid line}).
}
\end{figure}

The final question of resolution is whether, before introducing our
particles, we have enough resolution to model the turbulent flow field
correctly. This question applies both to the $N=512$ run into which we 
insert our particles, and to the $N=256$ runs we use to estimate the
mean and median accretion rates. To check convergence, we compare the
distribution of accretion rates we find from the $N=256$ simulations
at $\calm_0=5.0$ to the $N=512$ run. We do this for $\rb/\ell=0.20$, the
value for our particles. Figure \ref{convpdf} shows the PDFs. The
$N=512$ run mostly falls within the range of variation of the $N=256$
runs, although it is somewhat steeper, with less of the volume at high 
accretion rates. The average of the mean accretion rates in the
$N=256$ runs is $1.26\,\mdotzero$, and the average of the medians is
$0.36\,\mdotzero$. In comparison, for the $N=512$ run, the mean is
$0.90\,\mdotzero$, and the median is $0.34\,\mdotzero$. The agreement
on the medians is very good, and well within the variation among the
$N=256$ runs, but the agreement on the means is only
fair. The difference in quality of convergence probably occurs because 
the medians are sensitive to the peak of the distribution, but the
means depend more on the low vorticity, high accretion rate tail of
the distribution that is more sensitive to resolution. Thus, our means
probably have a $\sim 25\%$ systematic error.

\begin{figure}
\epsfig{file=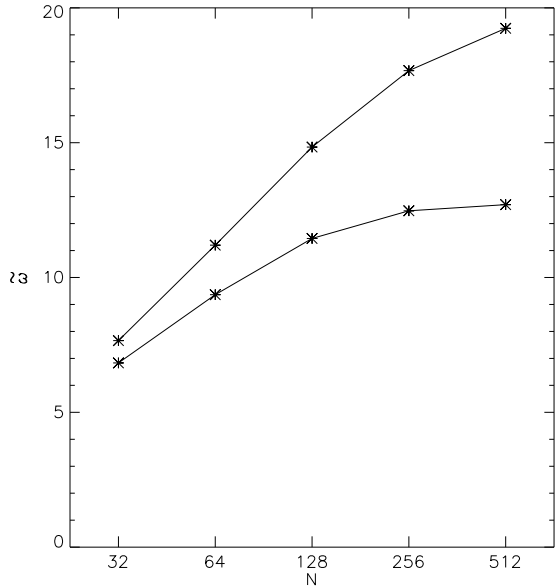}
\caption{\label{vortsize}
Mean (\textit{upper line}) and median (\textit{lower line}) normalized 
vorticity $\tilde{\omega}$ versus resolution.
}
\end{figure}

As a final note, the behavior of vorticity with resolution is
interesting, and helps explain why we are better converged in the
median than in the mean accretion rate and why the mean is decreasing
with resolution. We compute the mean and median
vorticity in the velocity field of our $N=512$ run when $\calm_0=5.0$,
just before we insert the
sink particles. We then sample the simulation down to $N=256$,
$N=128$, $N=64$, and $N=32$ and compute the mean and median vorticity
for those resolutions. Both the mean and median vorticity increase
monotonically with $N$, but the median vorticity appears to be
converging while the mean only shows slight curvature and may actually 
be diverging. We show this in Figure \ref{vortsize}.
The mean is controlled by a high vorticity tail, and it appears 
likely that there is no converged value for it until one reaches the
dissipation scale, either due to grid viscosity (in a simulation) or
due to physical viscosity (in a real fluid). In contrast, the median
appears to be well-defined and converged at $N=512$; the difference in 
median between the $N=256$ and $N=512$ resolutions is less than
$2\%$, and the median vorticity is rising with $N$ only as
$N^{0.026}$. A similar trend is visible in Figure \ref{vortpdfplot},
where the $N=512$ run peaks at roughly the same location as the
$N=256$ runs, but has a longer high-vorticity tail.

The divergence of the mean does not affect the
convergence of our results, however. The high vorticity tail that
causes the mean vorticity to diverge corresponds to low accretion
rates. Thus, it will not cause the mean accretion rate to diverge. Nor 
will it cause our accretion rate to go to zero as resolution
increases. The high vorticity tail will simply contribute less and
less to the mean accretion rate as the resolution increases, but the
contribution from vorticities near the median will not decrease
because the median does not change with resolution. Thus, we can
regard our $\sim 25\%$ change between the $N=256$ and $N=512$ runs as a
rough estimate of the likely error from imperfect
convergence. However, we conjecture that vorticity on scales much
smaller than $\rbh$ will not significantly affect the accretion
rate. In our $N=512$ runs, we have $\rbh/\Delta x=4$ on
the base AMR level. (Once we insert the particles we
greatly increase the resolution around them, but our estimates of
$\phimean$ are based on the simulation cube before that insertion).
Since we already have marginal resolution at $N=512$, we believe it is
unlikely that increasing resolution past $N=512$ will stronlgy affect
$\phimean$. Also
note that this convergence issue only affects accretion when
$\rb/\ell$ is large enough to affect the accretion rate. The velocity
PDFs, as shown in Figure \ref{velpdfplot}, are already very well
converged at $N=256$, so only when vorticity becomes important are our 
results uncertain.

\subsection{Limitations}
\label{limitations}

There are several limitations to our results. First, we have neglected 
magnetic fields. These are likely to suppress the accretion rate
relative to what we have found, because they will restrict the flow of 
gas across magnetic field lines to an accreting object. In molecular
clouds, where the Alfv\'en Mach number is of order unity
\citep{mckee89, mckee93, crutcher99} or less \citep{padoan04b}, the
effect is likely to be of order unity. However, magnetohydrodynamic
simulations similar to the ones we have performed will be required to
make a more definitive statement. A related limitation is that we have 
not considered a self-gravitating medium. If the gas is strongly
self-gravitating, its density and velocity distributions will change,
and our results for $\phimean$ and $\phimed$ may need to be
modified. However, our central result that vorticity is critical in
setting the accretion rate will not be affected by the addition of
self-gravity.

A second limitation is that our approach is likely to break down for
subsonic turbulence. The distribution of densities and velocities is
different in subsonically turbulent media than in supersonically
turbulent ones, so our results will not extrapolate to that
regime. For very subsonic turbulence where there is some vorticity
present, it is likely that the accretion flow approaches the
vorticity-dominated low $\oms$ regime identified by
\citet{krumholz05b}. The transition should occur for values of
$\calm_0$ near unity. However, we have not mapped out the details of
the change from one regime to the other, so we must be
careful in attempting to apply our theory to the warm atomic or
ionized phases of the ISM, since the motions in these phases are
generally transsonic or subsonic.

Third, our results are only applicable to objects with Bondi-Hoyle
radii significantly smaller than the size of the turbulent gas cloud
in which they are accreting. Our estimate of the accretion rate is
based on the assumption that, on large scales, the turbulent gas is
uniform and can be roughly modeled by a periodic box. However, if
$\rbh \gtsim \ell$, then the turbulent gas will become centrally
condensed due the gravity of the accretor, and our assumption of
uniformity will fail.

The final major limitation is that we have only considered the case in
which the accreting
object is not moving through the turbulent medium at a velocity
comparable to or larger than the velocity dispersion of the gas. In
such a situation an accretor should undergo velocity-dominated
accretion regardless of its position in the turbulent medium. The
accretion rate will be the standard Bondi-Hoyle rate, using the local
density as $\rho_{\infty}$. Consequently, the distribution of
accretion rates should be proportional to the distribution of
densities, which is well measured from simulations
\citep[e.g.][]{padoan99}. This limitation prevents us from applying
our theory to neutron stars or black holes passing through molecular
clouds, since these objects generally have velocity much larger than
the turbulent velocity dispersions of GMCs.

One might think that our assumption of isothermality also presents a
considerable limitation. However, this is not likely to be the
case. In ordinary Bondi-Hoyle accretion, changing the equation of
state only changes the accretion rate by factors of order unity
\citep{ruffert94a, ruffert94b}. Thus,
a different equation of state is not likely to change the nature of
the accretion flow onto a particle. The only way that a change in the
equation of state might affect our accretion rates is by changing the
distribution of densities, velocities, and vorticities in the
turbulent gas. Since we are assuming the turbulence is highly
supersonic, a different equation of state is unlikely to affect the
velocity field very much; thermal energy will always be considerably
less than kinetic energy over most of the flow. The remaining
potential effect of non-isothermality is on the density
field. \citet{passot98} show that isothermal turbulence leads to a PDF
of densities that is lognormal. Polytropic indices $\gamma<1$ lead
the PDF to develop a power-law tail at high densities, and indices
$\gamma>1$ lead to a power-law tail at low densities. The latter is
unlikely to affect the accretion rate, since the low density gas that
is affected does not contribute significantly to the mean or median
accretion rate. If $\gamma<1$, there will be more gas at high
densities than in the isothermal case, and the accretion rate is
likely to be larger than we have estimated by an amount that depends
on how small $\gamma$ is. Our results are therefore likely to change
only by order unity for $\gamma>1$, and will only change by much more
than that if $\gamma\ll 1$.

\subsection{Implications}
\label{implications}

Our results have implications for several astrophysical problems. Here 
we sketch out three. The first is the problem of pre-main sequence
accretion. \citet{padoan05} consider accretion onto
T Tauri stars from a turbulent medium using the
\textit{ansatz} $\mdotturb=\mdotbh$. (Note that this is slow accretion
that takes place after the pre-stellar core has collapsed and been
accreted, not the competitive accretion model of star formation. We
discuss competitive accretion below.) They find an accretion rate
consistent with observed accretion rates onto pre-main
sequence stars and brown dwarfs in nearby low-mass star-forming
regions. We have shown that using $\mdotturb=\mdotbh$
can lead to significant overestimates of the accretion
rate, and that the mean and median accretion rate are quite
different. Accretion rates are inferred from observations of the
accretion luminosity, which is sensitive to the accretion rate on time 
scales far smaller than the crossing time of the system. Thus, one
should use the median rather than the mean accretion rate when
comparing to individual accreting objects. However, we find that
Padoan et al.'s conclusions are not affected too strongly by our results,
provided their application is limited to the low-mass star-forming
regions to which Padoan et al. compare their models. Such regions have
widely distributed pre-main sequence stars and star-forming clumps
with large radii. If stars are forming in a turbulent clump of size
$\ell\sim 5$ pc, as Padoan et al. assume, then for a 1 $\msun$ star
$\rb/\ell\approx 0.02$ (for 10 K gas), and vorticity is relatively
unimportant. 

On the other hand, massive star-forming regions are more compact and
turbulent, and star-forming clumps there are more typically on scales
$\ell\approx 0.5$ pc, with Mach numbers $\calm_0\sim 20$
\citep{plume97}. Levels of turbulence are not significantly lower in
starless cores, so it is likely that these Mach numbers are typical
even at early stages of the star formation process
\citep{yonekura05}. For such a region $\rb/\ell\approx 0.2$ for a 1
$\msun$ star, and vorticity can suppress accretion. This
conclusion leads to the possibility of an interesting observational
test of our conclusions. Assuming that Padoan et al. are correct and
that accretion onto protostars comes primarily from accretion of
gas from the environment (rather than from some internal process
involving an accretion disk), one could bin accreting stars by mass,
and look for a change in the dispersion of accretion rates versus
mass. This break should occur because the mean and median accretion
rate begin to be affected by vorticity at different values of
$\rb/\ell$. Using (\ref{phimeanfit}), in a clump with $\calm_0=20$,
$\phimean$ is independent of mass for stars of mass $M<1$ $\msun$,
and $\phimean \propto M^{-0.7}$ for more massive stars. On the other
hand, $\phimed$ is independent of mass up to a mass of
$130$ $\msun$, vastly larger than any observed protostar. Thus, above
$1$ $\msun$ the ratio $\phimean/\phimed\propto M^{-0.7}$, and the
dispersion of accretion rates, which depends on this ratio, should
decrease with mass. Given a reasonably large sample of accreting
protostars in a dense clump with a range of masses above and below $1$
$\msun$, one could see this decrease in dispersion with mass. One
could also in principle look for the change in $\phimean$ with mass
directly, but this is more difficult because one would need a very
large sample of accreting stars to determine $\phimean$
accurately. Measuring the dispersion requires fewer sources. Even so,
measuring accretion luminosities of T Tauri stars in massive star
forming regions is a significant observational challenge due to
distance, confusion, and extinction. Unfortunately one cannot test our 
model using more massive stars, because the rate of Bondi-Hoyle
accretion onto stars larger than $\sim 10$ $\msun$ is likely to be
substantially reduced by radiaton pressure \citep{edgar04}.

A second potential implication is in the area of competitive
accretion. In this model of star formation, the initial mass function
is determined by seed protostars accreting unbound gas within the
turbulent molecular clump where they form \citep[e.g.][]{bonnell97,
klessen00a, klessen01, bonnell01a, bonnell01b, bonnell04}. Our theory
does not apply during the first stage of competitive accretion, when a
star is still accreting from its initial bound core, for two
reasons. First, the core is bound to the star and to itself, while we
have considered accretion of unbound gas. Second, the star is
co-moving with the gas in its core, so it is not randomly sampling the
turbulent flow. Once the star accretes its parent core, these limits
no longer apply and one may use our theory. The star is no longer
accreting bound gas, and, since the time required to accrete the bound
core \citep[which is comparable to the dynamical time of the parent
clump --][]{mckee03} will allow the star to become dynamically
decoupled from the turbulent flow, the star will be randomly sampling
the gas in the molecular clump to which it is gravitationally
confined.

Our results suggest that in this stage the accretion rate in
such clumps may be considerably different than a naive application of
the Bondi-Hoyle formula suggests. Simulations of competitive accretion
performed to date likely have too little resolution to accurately
model the effect we describe here. In a typical massive star-forming
clump, the velocity dispersion is $\sim 4$ km s$^{-1}$
\citep{plume97,yonekura05}, giving a Bondi-Hoyle radius for a $0.1$
$\msun$ seed protostar of only 5.5 AU, far smaller than the resolution
of the simulations performed thus far. Published calculations of
competitive accretion to date have largely avoided the problem by
using the lower velocity dispersions characteristic of low-mass star
forming regions. However, it represents a formidable obstacle to
simulating the formation of rich clusters from massive clumps. It is
unclear what effect failing to resolve this radius would have on the
computed accretion rate. Competitive accretion simulations generally
use smoothed-particle hydrodynamics with sink particles, and these
methods compare the kinetic and potential energies of a gas particle
in an attempt to determine whether it is bound to a sink particle
before accreting it \citep{bate95}. As a result, the most likely error
is an understimate of the accretion rate caused by failure to resolve
the Bondi-Hoyle accretion shock cone and the resulting loss of gas
kinetic energy.

A third potential implication of our result is for the accretion rates
of black holes in the centers of galaxies. Some simulations and
semi-analytic models of the growth of galactic center BHs assume that
the black holes accrete at the Bondi rate
\citep[e.g.][]{springel05}. If the gas around the black hole is
turbulent, however, using the Bondi rate may lead to significant
errors. It is uncertain whether the accretion process onto galactic
center BHs can be described by Bondi accretion at all, but, if it can,
then simulations and analytic models should use the turbulent
accretion rate we have found rather than the unmodified Bondi rate. 

\section{Conclusions}
\label{bhtconclusion}

The primary result of our investigation is that the accretion rate in
a turbulent medium is significantly enhanced over the zeroth-order
estimate one obtains by simply using the turbulent velocity dispersion 
in the Bondi-Hoyle formula. However, this enhancement may be partially 
or completely offset by the effects of vorticity within the turbulent
medium, and may even be suppressed below the Bondi-Hoyle rate for
Mach numbers near or below unity. For this reason, simply using the
Bondi-Hoyle formula either by treating the global velocity dispersion
as a sound speed or by applying the formula to every point within the
medium \citep{padoan05}
does not produce the correct accretion rate. On the other
hand, approximating the accretion process as being dominated by either
velocity or vorticity, and using the lesser of the two implied
accretion rates, appears very consistent with the results of simulations.

We use this simple model to compute the probability distribution of
accretion rates onto point particles in a turbulent medium. The PDF
of accretion rates is lognormal, and we give simple formulae that
specify the shape of the lognormal distribution, and the mean and
median accretion rates, over a relatively wide range of
accretor masses and turbulent Mach numbers. The mean and median
accretion rates can be quite different, due to the long tail of the
lognormal accretion rate distribution, and it takes an accreting
particle of order a crossing time of the turbulent region before its
time-averaged accretion rate approaches the mean for the region.

Our findings suggest that a number of previously published numerical
and analytical results may need to be reconsidered, and that
future work may have to meet quite severe resolution
constraints. Estimates of accretion rates in star-forming regions
using the simple Bondi-Hoyle formula are probably incorrect, and
simulations of competitive accretion that do not resolve the
Bondi-Hoyle radius, and possibly even smaller scales, are likely to
produce incorrect accretion rates. Simulations of star formation that
have been completed to date have largely avoided this problem by
not simulating the very large velocity dispersions found in massive
star-forming regions, and our work shows that expanding the
simulations to these regions will require much higher resolution.
We leave a more detailed reconsideration of these problems, using our
newly developed theory of Bondi-Hoyle accretion in a turbulent medium,
to future work.

\acknowledgements The authors thank Phil Marcus and Eliot Quatert for
useful discussions, and the referee for useful comments. This work was
supported by: NASA through Hubble
Fellowship grant \#HSF-HF-01186 awarded by the Space Telescope Science
Institute, which is operated by the Association of Universities for
Research in Astronomy, Inc., for NASA, under contract NAS 5-26555
(MRK); NASA GSRP grant NGT 2-52278 (MRK); NSF grant AST-0098365 (CFM);
NASA ATP grant NAG 5-12042 (RIK and CFM); and the US Department of
Energy at the Lawrence Livermore National Laboratory under contract
W-7405-Eng-48 (RIK and MRK). This research used computational
resources of the National Energy Research Scientific Computing Center,
which is supported by the Office of Science of the U.S. Department of
Energy under Contract No. DE-AC03-76SF00098, through ERCAP grant
80325; the NSF San Diego Supercomputer Center through NPACI program
grant UCB267; and the US Department of Energy at the Lawrence
Livermore National Laboratory under contract W-7405-Eng-48.

\bibliographystyle{apj}
\bibliography{ms}

\end{document}